\documentstyle[11pt,aaspp4]{article}

\begin{document}


\def\ast{\mathchar"2203} \mathcode`*="002A 
\def\larrow{\leftarrow} \def\rarrow{\rightarrow} 
\def\=={{\equiv}} 
\def\vinf{v_{\infty}} 
\def\solar{\odot} 
\def\Mdot{\dot M} 
\def\mdot{\dot M} 
\def\zp{{z^{\prime}}} 
\def\rp{{r^{\prime}}} 
\def\Lbol{{L_{bol}}} 
\def\Ro{{R_{o}}} 
\def\Rstar{{R_\ast}} 
\def\rosat{{\it ROSAT}}
\def\bbxrt{{\it BBXRT}}
\def \asca{{\it ASCA}}
\def \axaf{{\it AXAF}}
\def \zpup{\hbox{$\zeta$ Pup}}
\def \fx{\relax \ifmmode {f_{\rm X}} \else $f_{\rm X}$\fi}
\def \Lx{\relax \ifmmode {L_{\rm X}} \else $L_{\rm X}$\fi}
\def \lx{\relax \ifmmode {L_{\rm X}} \else $L_{\rm X}$\fi}
\def \lxlb{\relax \ifmmode {L_{\rm X}/L_{\rm Bol}} \else $L_{\rm
    X}/L_{\rm Bol}$\fi} 
\def \Lbol{\relax\ifmmode{L_{\rm Bol}}\else $L_{\rm Bol}$\fi}
\def \lb{\relax\ifmmode{L_{\rm Bol}}\else $L_{\rm Bol}$\fi}

\def\Msunyr{\hbox{${\rm M}_\odot\,$  yr $^{-1}$}}
\def\msunyr{\hbox{${\rm M}_\odot\,$  yr $^{-1}$}}
\def \taunu{\relax \ifmmode {\tau_{\nu}} \else $\tau_{\nu}$ \fi}
\def \sigmanu{\relax \ifmmode {\sigma_{\nu}} \else $\sigma_{\nu}$ \fi}
\def \ecma{\hbox{$\epsilon$ CMa}}
\def \einstein{{\it Einstein}}
\def \pspc{{\rm PSPC}}
\def \euve{{\it EUVE}}
\def \cf{{\it c.f.}}
\def \eg{{\it e.g.}}
\def \etal{et~al.}
\def \ie{{\it i.e.}}
\def \mdovervinf{\relax \ifmmode {\dot M}/v_{\infty} \else ${\dot M}/v_{\infty}$\fi}
\def\Exp{{e}} 
\def\ltwig{\mathrel{\spose{\lower 3pt\hbox{$\mathchar"218$}} 
     \raise 2.0pt\hbox{$\mathchar"13C$}}} 
\def\gtwig{\mathrel{\spose{\lower 3pt\hbox{$\mathchar"218$}}  
     \raise 2.0pt\hbox{$\mathchar"13E$}}} 
\def\spose#1{\hbox to 0pt{#1\hss}} 
\def\blankline{\par\vskip \baselineskip}


\title{A Simple Scaling Analysis of X-ray Emission and
Absorption in Hot-Star Winds}

\author{Stanley P. Owocki and David H. Cohen}
\affil{Bartol Research Institute, University of Delaware,
  Newark, Delaware, 19716. E-mail:  owocki@bartol.udel.edu, cohen@bartol.udel.edu}

\begin{abstract}

We present a simple analysis of X-ray emission and absorption for hot-star winds,
designed to explore the natural scalings of the observed X-ray luminosity with wind and
stellar  properties.   
We show that an exospheric approximation, in which all of the emission  above the optical
depth unity radius escapes the wind, reproduces very well the formal solution for
radiation transport through a spherically symmetric wind.   
Using this approximation we find that the X-ray luminosity $\lx$ scales naturally with 
the wind density parameter $\Mdot/\vinf$, obtaining 
$\lx \sim (\Mdot/\vinf)^2$ for optically thin winds, and 
$\lx \sim (\Mdot/\vinf)^{1+s}$ for optically thick winds with an X-ray filling factor
that varies in radius as $f \sim r^s$.
These scalings with wind density contrast with the commonly inferred empirical
scalings of X-ray luminosity $\lx$ with bolometric luminosity $\lb$.
The empirically derived linear scaling of $\lx \sim \lb$ for thick winds 
can however be reproduced, through a delicate
cancellation of emission and absorption, if one assumes modest radial fall-off in 
the X-ray filling factor ($s \approx -0.25$ or $s \approx -0.4$, depending on details 
of the secondary scaling of wind density with luminosity).
We also explore the nature of the X-ray spectral energy distribution in the context
of this model, and find that the spectrum is divided into a soft, optically thick
part and a hard, optically thin part.  
Finally, we conclude that the energy-dependent emissivity must have a high-energy
cut-off, corresponding to the maximum shock energy, in order to reproduce the
general trends seen in X-ray spectral energy distributions of hot stars. 

\vspace{0.1in}

\end{abstract}

\section{Introduction}

One of the great surprises of the {\it Einstein} X-ray satellite
mission was the observation that hot, luminous, OB stars are
strong emitters of soft ($\sim 1$ keV) X-rays (Harnden et al. 1979;
Seward et al. 1979).
Unlike the cooler, late-type stars, for which the X-ray emission
was found to scale with the star's rotation, the observed X-rays from hot
stars show a roughly linear proportionality to the stellar
bolometric luminosity, $L_x \approx 10^{-7} L_{bol}$
(Long \& White 1980; Pallavacini et al. 1981; Chlebowski, Harnden,
and Sciortino 1989).
Observations from subsequent X-ray satellites, most notably \rosat,
have generally confirmed this earlier scaling result, with some minor
refinements.
For example, from a \rosat\ survey of 42 O stars, Kudritzki \etal\ (1996) find
$\log \lxlb \approx -6.7 \pm 0.35$,
with a somewhat tighter relationship when a weak dependence on
a characterisic wind density \mdovervinf\ is included,
$\lx \propto (\mdovervinf)^{-0.38} \lb^{1.34}$.

Even before these detections, X-ray emission from a narrow stellar
``corona'' was postulated to explain the superionization seen in the
UV spectra of OB stars (Cassinelli \& Olson 1979).
This coronal model was further refined by Waldron (1984), but other
studies have limited the extent, temperature, and fractional contribution
of the total X-ray output of such purported O-star coronae
(Cassinelli, Olson, \& Stalio 1978; Nordsieck, Cassinelli, \& Anderson 1981;
Cassinelli \& Swank 1983; Baade \& Lucy 1987; MacFarlane et al. 1993).
A much more generally favored scenario has been that these X-rays
arise from {\it shocks} that form in the wind from strong intrinsic instabilities in
the line-driving mechanism 
(Lucy \& Solomon 1970; MacGregor, Hartmann, \& Raymond 1979;
Owocki \& Rybicki 1984, 1985).
Lucy \& White (1980) and Lucy (1982) proposed phenomenological
models in which wind material accelerated by this line-driven
instability rams into ambient, ``shadowed'' material, forming
a {\it forward} shock that accelerates and heats that material,
producing X-rays.

Dynamical simulation models of the nonlinear evolution
of this instability have since indicated a quite different wind structure,
dominated by {\it reverse} shocks that {\it decelerate} very rarefied,
instability-accelerated material as it rams into compressed,
dense shells ahead of it (Owocki, Castor, \& Rybicki 1988).
Since only a small fraction of wind material is actually heated by
passage through such reverse shocks, the resulting direct X-ray emission
is very low, generally well below the observed value (Feldmeier et al. 1997a).
To overcome this limitation, subsequent wind structure simulations by
Feldmeier et al. (1997b) introduced a chaotic, turbulent perturbation
at the photospheric lower boundary.  This seeds a wind structure
with substantial velocity dispersion among the dense shells, which
thus collide amongst themselves, yielding much more shock-heated
material. For the specific case of a massive OB supergiant
wind with rather strong and chaotic base perturbations,
the X-ray emission produced in such models can be made to
approach the observed values.
However, the dynamical simulations for particular cases have
thus far made no attempt to explain the broad scaling of observed
X-ray emission with stellar luminosity.

The  goal of the present paper is to provide a firm basis for
understanding these more general scaling properties of hot-star X-rays.
Our approach here eschews detailed dynamical simulations, and instead
examines phenomenalogical models of the X-ray emission, in
order to define more clearly what overall properties
are needed from instability-generated wind structure simulations
to produce these observed scaling relations.
A major point of our analysis is that a {\it linear} scaling
of {\it observed} $L_x$ with $L_{bol}$ does not arise naturally
from expected properties of a wind-shock model, but
requires a rather delicate balance of X-ray absorption and
emission, which in turn requires a special form for
the radial distribution of wind shocks.
A further point is that such a balance between emission and
absorption should be associated with specific trends in
broad-band X-ray spectral properties.

The remainder of this paper is organized as follows.
Beginning from the formal solution to the radiation transfer equation
for X-rays emitted throughout an expanding stellar wind (\S 2.1),
we develop a simple ``exospheric'' approximation through which
we derive a general scaling law for X-ray luminosity from
a simple power-law emission model (\S 2.2). By relating this
to wind and stellar scaling laws we identify parameter requirements
for reproducing the observationally inferred linear dependence of $L_x$
on $L_{bol}$ (\S 2.3).  We  next examine the energy dependence
of the X-ray spectrum (\S 3.1) and then integrate the expression for the X-ray spectrum
over energy in order to verify the result from the previous section that gave the linear
$\lx \sim \lb$ scaling (\S 3.2).   The final section (\S 4) summarizes our conclusions.
In two appendices we examine the effect of assuming a constant wind velocity and verify the 
validity of the assumption of a smooth X-ray emissivity in
terms of detailed numerical plasma emission models.

\section{Radiation Transfer Through an Expanding, Spherical Stellar
  Wind} \label{sec:rt}

\subsection{Formal Solution for Constant Velocity Expansion}
\label{subsec:constvel}

To develop a simplified description of X-ray emission and absorption in a spherically
symmetric, expanding envelope, we initially consider monochromatic X-rays, deferring the 
discussion of energy dependence to the next section.  Let us
first define the usual
$(p,z)$ coordinates such that $z$ is a coordinate along a ray, and $p$ is the `impact
parameter', or minimum radial distance of the ray from the origin.  From a formal
solution of the transfer equation for X-rays emitted at a radius $r=\sqrt{p^2 + z^2}$
with emissivity $\eta (r)$, the specific intensity observed at large radii ($z
\rightarrow \infty$) is
\begin{equation}
I_p = \int_{-\infty}^\infty \eta (r) \Exp^{-\tau(p,z)} \,  dz .
\label{eqn:ip}
\end{equation}
The total X-ray luminosity is then given by integration over all rays
\begin{equation}
L_x = 8 \pi^2 \int_0^\infty I_p \, p \, dp.
\label{eqn:lx}
\end{equation}
Here the ray optical depth is of the form
\begin{equation}
\tau (p,z) = \int_z^\infty \kappa_x (r(p,\zp)) \, \rho (r(p,\zp)) \, d \zp ,
\label{eqn:tau}
\end{equation}
where $\rho$ is the mass density, and $\kappa_x$ is the X-ray
absorption cross section per unit mass. We note that the opacity arises in relatively cool, 
unshocked portions of the wind, and is due primarily to K-shell bound-free transitions in 
He, C, N, and O, Ne, Na, Mg, and Si.

For an instability-generated shock model, X-ray emission is only
expected once the wind has reached a substantial fraction ($\gtwig$50\%)
of the terminal speed $\vinf$, implying that the radial variation
of mean wind density is within an order unity factor of that for a
{\it constant} velocity (Owocki 1994; Feldmeier 1995). For simplicity, let
us thus consider the case of expansion at a constant velocity
$\vinf$ (but see Appendix \ref{App:beta} for a treatment of a  beta velocity law).
Moreover, since ionization fractions are also fairly constant beyond
about $0.5\vinf$ (MacFarlane, Cohen, \& Wang 1994), we assume also a radially
constant X-ray mass-absorption coefficient, $\kappa_x$.
The radial optical depth is then given simply by
\begin{equation}
\tau(0,r) =
{\kappa_x \Mdot \over 4 \pi \vinf } \int_r^\infty { dr' \over r'^2}
=
{ R_1 \over r } ,
\label{eqn:taur}
\end{equation}
where $\Mdot$ is the mass loss rate, and
\begin{equation}
R_1 \== { \kappa_x \Mdot \over 4 \pi \vinf }
\label{r1def}
\end{equation}
defines the radius of unit radial optical depth,
i.e.  $\tau (0,R_1) \== 1$.
For nonradial ($p \neq 0$) rays, we obtain
\begin{equation}
\tau (p,z) =
R_1 \int_z^\infty {d\zp \over p^2 + \zp^2 }
 =  { R_1 \over p } \arccos { \left [ z \over \sqrt{p^2+z^2} \right ] }
 = { R_1 \over p } \theta (p,z),
\label{eqn:tau2}
\end{equation}
where $\theta$ is the angle between the $+z$ ray direction and
an outward radius vector.
This suggests it would be convenient to recast the intensity integral
(\ref{eqn:ip}) in terms of this angle,
\begin{equation}
I_p = p \int_0^\pi {\eta (r(p,\theta)) \over
                   \sin^2 \theta \, \Exp^{R_1 \theta /p} }
                     \,  d\theta .
\label{eqn:ip_of_theta}
\end{equation}

To proceed, we next note that X-ray emission arises primarily from two-body
processes -- recombination and collisional excitation --
and thus has a volume emissivity $\eta ~ {\rm (ergs~s^{-1}~cm^{-3}~ster^{-1})}$ that is
proportional to the square of the density,
\begin{mathletters}
\begin{eqnarray}
\eta (r) = \xi C_s f_m \rho^2 (r)
\label{eqn:eta}
\\
= \xi C_s ^2f_v \rho^2 (r),
\label{eqn:eta_fv}
\end{eqnarray}
\end{mathletters}
where $f_m$ is the {\it mass} fraction, or mass filling factor, of ambient wind that is
X-ray emitting, and
$C_s\== \rho_s/\rho$ is a factor to correct for any differences in the density
$\rho_s$ of shocked material relative to the ambient wind density $\rho$ (which assumes
a smooth wind, given by $\rho \== \Mdot / 4\pi r^2v$).  We note that the factor $C_s$
accounts both for the post-shock compression and the deviations of the pre-shock
density from the ambient wind density.  The mass filling factor, $f_m$, is related to
the volume filling factor by $f_v = f_m/C_s$. The emission coefficient $\xi$ (${\rm
ergs~cm^3~s^{-1}~g^{-2}}$) is related to the commonly used cooling function $\Lambda$
(ergs s$^{-1}$ cm$^3$) (e.g., Raymond \& Smith 1977) by $\xi = \Lambda/4 \pi \mu_e
\mu_p$, where $\mu_e$ and $\mu_p$ are the mean mass per electron and proton,
respectively. Under typical stellar wind conditions $\xi$ is independent of density (at
least when considered at the coarse spectral resolution associated with current X-ray
detectors), and depends mainly on the electron temperature in the post-shock region.
Assuming for now that the factor $\xi C_s^2 f_v$ is spatially constant, the intensity
integral (\ref{eqn:ip_of_theta}) can be rewritten and evaluated as
\begin{mathletters}
\begin{eqnarray}
  I_p = {\xi C_s^2 f_v \over p^3} \left ( {R_1 \over \kappa_x } \right )^2
\int_0^\pi \sin^2 \theta \, \Exp^{-R_1 \theta /p} \, d\theta
\\
  = 2 \xi C_s^2 f_v { R_1 \over \kappa_x^2 } \left ( { 1 - \Exp^{-\pi
        R_1 /p} \over R_1^2 + 4 p^2 } \right ) .
\end{eqnarray}
\label{eqn:ip_of_theta2}
\end{mathletters}
Application to the luminosity integral (\ref{eqn:lx}) then yields
\begin{mathletters}
\begin{eqnarray}
L_x = 4\pi^2 \xi C_s^2 f_v { R_1 \over \kappa_x^2 } 
\int_0^\infty{ 1 - \Exp^{-2 \pi x} \over x ( 1 + x^2 ) } \,  dx
\\
     \approx 9.76 \pi^2 \xi C_s^2 f_v { R_1 \over \kappa_x^2 }
\\
 =  2.44 \pi {\xi C_s^2 f_v \over \kappa_x } \, {\Mdot \over \vinf } ,
\end{eqnarray}
\label{eqn:lx2}
\end{mathletters}
where the approximate equality follows from direct numerical evaluation of the
integral.  Note that, although the wind volume emission varies as the density squared,
the X-ray luminosity escaping the wind scales only {\it linearly} with the density
parameter $\Mdot/\vinf$, due to the effect of wind attenuation.

\subsection{Exospheric Approximation for Power-Law Radial Emission}
\label{sec:exo-powerlaw}

It is of interest to compare this numerical result with the heuristic
formula from an ``exospheric approximation'' (Cohen \etal\  1996)
that estimates the X-ray luminosity from the volume integral of the
outward (\ie\/ radiated into $2\pi$ ster) emission beyond the radius with 
unit radial optical depth,
\begin{equation}
L_x \approx 8 \pi^2 \int_{R_1}^\infty  \eta (r) \, r^2 dr 
= {2 \pi \xi C_s^2 f_v \Mdot  \over \kappa_x \vinf} ,
\label{eqn:exospheric}
\end{equation}
where the latter equality follows from straightforward analytic integration.  The  rough
agreement with the numerical result (eq. [\ref{eqn:lx2}]) suggests that this simple
exospheric formula should be quite useful for estimating $L_x$ in more complicated cases
for which the required integrals are difficult to evaluate.

Let us thus next estimate $L_x$ for the somewhat more general case
when the factor $\xi C_s^2 f_v$ varies as
a power-law beyond some lower boundary radius $\Ro$ for
X-ray emission,
\begin{mathletters}
\begin{eqnarray}
\eta (r) = \rho^2 \, (\xi C_s^2 f_v)_o
\left ( { r \over \Ro } \right )^{s} ~~ ; ~ r > \Ro ~
\label{eqn:radialvar}
\\
         = 0  ~~~~~~~~~~~~~~~~~~~~~~~~~~ ; ~ r < \Ro ,
\end{eqnarray}
\end{mathletters}
where $(\xi C_s^2 f_v)_o$ is the emission factor at $r=\Ro$.
Unlike for the special case
$\Ro=R_{\ast}$ and $s=0$ considered  above,
exact evaluation for $L_x$ would now generally require numerical evaluation
of a double integral over $p$ and $z$.
However, application of the exospheric formula
(eq. [\ref{eqn:exospheric}])
readily yields an approximate analytic scaling law for the X-ray luminosity,
\begin{equation}
L_x \approx { (\xi C_s^2 f_v)_o \over 2(1 - s )} \left ( { \Mdot \over \vinf }
\right )^2\Ro^{-s} (\max[\Ro,R_1])^{s-1}.
\label{eqn:Lx_of_s}
\end{equation}
We note
parenthetically that this
can be recast as a fitting formula,
\begin{equation}
L_x \approx {(\Lambda C_s^2 f_v)_o \over 2 (1 - s)}
EM_w \left ( {\Ro \over \max(\Ro,R_1)}
\right ) ^{1-s} ,
\label{eqn:Lxfitting}
\end{equation}
where $EM_w = \int_\Ro^\infty n^2 d{Vol} $ is the  total wind
emission measure above $\Ro$.  Alternatively, we can express this
in terms of the basic stellar wind parameters,
\begin{mathletters}
\begin{eqnarray}
L_x \approx { (\xi C_s^2 f_v)_o \over 2(1 - s ) R_o} \,  
\left ( {4 \pi \Ro \over \kappa_x } \right )^{1-s} \,
\, \left ( { \Mdot \over \vinf } \right )^{1+s} ~~~~~ ; ~ \Ro < R_1
{\rm ~(optically~thick)}
\\
\label{eqn:lx_vs_stellarparams_b}
~~~~ \approx { (\xi C_s^2 f_v)_o \over 2 (1-s) R_o } \left ( { \Mdot \over
    \vinf } \right )^2 ~~~~~~~~~~~~~~~~~~~~~~~~~~ ; ~ \Ro > R_1 {\rm
  ~(optically~thin)}.
\label{eqn:lx_vs_stellarparams_c}
\end{eqnarray}
\end{mathletters}
Equations (\ref{eqn:lx_vs_stellarparams_b}) and (\ref{eqn:lx_vs_stellarparams_c})
represent the X-ray luminosity for, respectively, the `thick' and `thin' wind cases. 
Note that the luminosity now scales with $(\Mdot/\vinf)^2$ for the thin wind, but only
with $(\Mdot/\vinf)^{1+s}$ for the thick wind.

We reiterate that these results have all been derived for monochromatic X-rays.  
We will see in \S 3 that a given stellar wind can be optically thick and some 
energies and optically thin at others, leading to spectral structure based on the 
energy-dependence of the X-ray attenuation.

\subsection{Relation to Wind Scaling Laws}
\label{sec:windscale}

The above simple analysis illustrates that the most direct scaling of the X-ray 
luminosity should be with wind density, and not, as has generally been inferred 
empirically, with the bolometric luminosity.
Let us thus next consider whether this inferred scaling could be explained through
secondary scaling of wind density with luminosity.
Within the standard Castor, Abbott, and Klein (1975; hereafter CAK) theory for
line-driven winds, the mass loss is predicted  to scale as
\begin{equation}
\Mdot \sim \lb^{1/\alpha'} \, M_{eff}^{1-1/\alpha'} ,
\label{mdcak}
\end{equation}
where $M_{eff}$ is the effective stellar mass ($M_{eff} = M_{\ast} (1 - \Gamma_{Edd})$), and
$\alpha' = \alpha - \delta$ is a combination of the two line parameters, 
$\alpha$ and $\delta$ defined by CAK and Abbott (1982), 
which respectively describe the distribution of line opacity and its dependence on
ionization level.

As first noted by Kudritzki, Lennon, and Puls (1995), this theoretical scaling is in good
general accord with an empirically derived `wind-momentum-lumosity' relation
\begin{equation}
\Mdot \vinf R_{\ast}^{1/2} ~ \sim ~ \lb^{1/\alpha'} ,
\label{eqn:momlum}
\end{equation}
wherein the connection with the CAK mass loss formula (\ref{mdcak}) follows from
the scaling of the wind terminal
speed with the surface escape speed,
\begin{equation}
\vinf  \sim \sqrt{ M_{eff} / R_{\ast}} ,
\label{eqn:vive}
\end{equation}
which is a well known theoretical (CAK; Pauldrach, Puls, \& Kudritzki 1986) and observational 
(e.g., Lamers, Snow, \& Lindholm 1995) result.
For O supergiants, the extensive analysis by Puls et al. (1996) empirically obtained a 
wind-momentum-luminosity relation with a slope implying
$\alpha' = 0.57 $, which thus gives 
\begin{equation}
{\Mdot \over \vinf } ~ \sim ~ { \lb^{1.75} \, R_{\ast}^{0.5} \over M_{\ast} (1 - \Gamma_{Edd}) } .
\label{eqn:mdov_vs_lrom}
\end{equation}

The dependence on the Eddington factor is significant only for stars with very high luminosities.  
Between an early-B star like $\lambda$ Sco and an O supergiant like \zpup, $(1 - \Gamma_{Edd})$ 
changes by only a factor of 2, while the bolometric luminosities vary by almost three orders 
of magnitude.  

To convert eq. (\ref{eqn:mdov_vs_lrom}) to a scaling with luminosity alone, we can ignore the 
weak dependence on stellar radius; but we may want to allow for a systematic 
trend between luminosity and mass.
For example, from stellar structure theory, one expects a quite strong trend
e.g. $\lb \sim  M_{\ast}^{2.5}$, yielding
${\Mdot / \vinf} \sim \lb^{1.35}$.
The observed $\lx \sim \lb$ relation would then be reproduced by the optically
thick wind scaling relation (\ref{eqn:lx_vs_stellarparams_b}) if $ s \approx -0.25 $.

By contrast, if we simply ignore any systematic trend between mass and luminosity,
the observed $\lx \sim \lb $ relation would be reproduced by a filling-factor index 
$s \approx -0.40$.
Either scaling is qualitatively consistent with detailed numerical simulations of the 
line-driven instability, which generally show a gradual drop-off in X-ray production 
with radius beyond the point where strong shocks begin to form in the wind. 
But overall it seems that, with such a wind-based model for X-ray emission,
connection with the observed $\lx \sim \lb $ is inherently indirect, requring
a rather specialized cancellation between the wind emission and absorption.

Finally, we note that for lower-density winds that are optically thin to X-rays,
the X-ray luminosity becomes independent of $s$, scaling as
\begin{equation}
\lx \sim \left ({\Mdot \over \vinf} \right)^2 \sim \lb^{2.7} ~ {\rm or} ~ \lb^{3.5},
\label{eqn:thick3.4}
\end{equation}
where the two scalings depend on whether one includes or
ignores the systematic variation of luminosity with mass.
Either form is in general accord with empirically inferred X-ray scalings, 
which show the dependence of X-rays on bolometric luminosity  becomes much 
steeper around early-B stars, corresponding roughly to where the winds are 
becoming optically thin (Cohen \etal\ 1997).

\section{Energy Dependence} \label{sec:sec_energy_dependence}
\subsection{Power-law Absorption and Emission} \label{sec:energy_dependence}

All of the above analysis is for monochromatic X-rays.  However, current
instruments (\einstein, \rosat, \asca) are sensitive to X-rays over almost two orders of
magnitude in energy, and even with their quite modest spectral resolutions, have
provided some basic information about the spectral energy distributions of X-rays in hot
stars.  So let us now consider the energy dependence of the absorption and emission,
and of the resulting luminosity spectrum.  Given the modest energy resolution of X-ray
observations to date, we focus here on the broad-band spectral properties. The X-ray
absorption -- apart from several prominent K-shell edges -- can be fit roughly by a
power-law in energy,
\begin{equation}
\kappa_x (E_x) \approx { \kappa_{xo} \over E_x^a }.
\label{eqn:kappa}
\end{equation}
Hillier et al. (1993) find, for $\zeta$ Pup (O4 If), $a \approx 2$ in
layers where helium remains doubly ionized, and $a \approx 2.8$ where it has
recombined to He$^+$.  For $\epsilon$ CMa (B2 II), detailed
calculations give $a \approx 1.8$ between 100 eV and 1 keV (Cohen et
al. 1996).
This energy dependence of opacity means a given wind can be optically thick at
some energies, and optically thin at others.
By setting $R_1 = \Ro$, we can derive a critical energy,
\begin{equation}
E_1 = \left ( { \kappa_{xo} \Mdot \over 4 \pi \vinf \Ro } \right )^{1/a} ,
\label{eqn:e1}
\end{equation}
that separates the parts of the spectrum for which the wind is optically thick
($E_x < E_1$) from those for which the wind is optically thin ($E_x > E_1$).

To derive the form of the luminosity spectrum one must specify the energy dependence of
emission as well as the absorption. For simplicity let us again assume that the
emissivity has a power-law energy dependence,
\begin{equation}
\xi (E_x) = { \xi_{\ast} \over E_x^b } ,
\label{eqn:xi_of_ex}
\end{equation}
which means that equation (\ref{eqn:radialvar}) can be written as
$ \eta = \rho^2 ( \xi_{\ast} C_s^2  f_v)_o E_x^{-b} (r/R_o)^s$.
In Appendix \ref{App:em} we show a comparison of this simple, smooth emissivity model to detailed
numerical simulations of collisional equilibrium plasmas. This comparison shows that at
the low resolution of current instruments, the line-dominated spectra are surprisingly
well approximated by a smooth function such as equation (\ref{eqn:xi_of_ex}), with a
power-law spectrum being produced by a power-law temperature distribution.  The
luminosity spectrum is then given by,
\begin{mathletters}
\begin{eqnarray} 
{ dL_x  \over dE_x } = L_1 \left ( { E_x \over E_1 } \right )^{a(1-s)-b} ; ~ E_x < E_1~
\label{eqn:lx_of_ex_a}
\\
        = L_1 \left ( { E_x \over E_1 } \right )^{-b} ~~~~~~~~ ; ~ E_x > E_1 ,
\label{eqn:lx_of_ex_b}
\end{eqnarray}
\end{mathletters}
where the luminosity at the energy at which wind optical depth is unity, $L_1 \==
L_x(E_1)$, has the form
\begin{equation}
L_1 =  { (\xi_{\ast} C_s^2 f_v)_o  \over 2 R_o ( 1 - s) } \,
\left ( {4 \pi \Ro \over \kappa_{xo} } \right )^{b/a}
         \left ( { \Mdot \over \vinf } \right )^{2-b/a} .
\label{eqn:lmax}
\end{equation}
While equations (\ref{eqn:lx_of_ex_a}) and (\ref{eqn:lx_of_ex_b}) are convenient for
plotting the spectrum for a given model, we can alternatively write this in a form that
more explicitly displays the scalings with stellar parameters,
\begin{mathletters}
\begin{eqnarray}
{ dL_x \over dE_x} = { (\xi_{\ast} C_s^2 f_v)_o  \over 2 R_o ( 1 - s) } \,
\left ( {4 \pi \Ro \over \kappa_{xo} } \right )^{1-s}
\, \left ( { \Mdot \over \vinf } \right )^{1+s} E_x^{a(1-s)-b} ~~ ; ~ E_x < E_1~
\label{eqn:lx_of_stellarpars_a}
\\
   ~~ =  { (\xi_{\ast} C_s^2 f_v)_o \over 2 \Ro (1-s) } \left (
   { \Mdot \over \vinf } \right )^2 E_x^{-b} 
~~~~~~~~~~~~~~~~~~~~~~~~~~~~; ~ E_x > E_1 ,
\label{eqn:lx_of_stellarpars_b}
\end{eqnarray}
\end{mathletters}
where, again, equations
(\ref{eqn:lx_of_stellarpars_a}) and
(\ref{eqn:lx_of_stellarpars_b}) represent the
expressions for, respectively, the optically thick
and optically thin wind domains
(cf. eqs. [\ref{eqn:lx_vs_stellarparams_b}] and [\ref{eqn:lx_vs_stellarparams_c}]).

\begin{figure}
\plotone{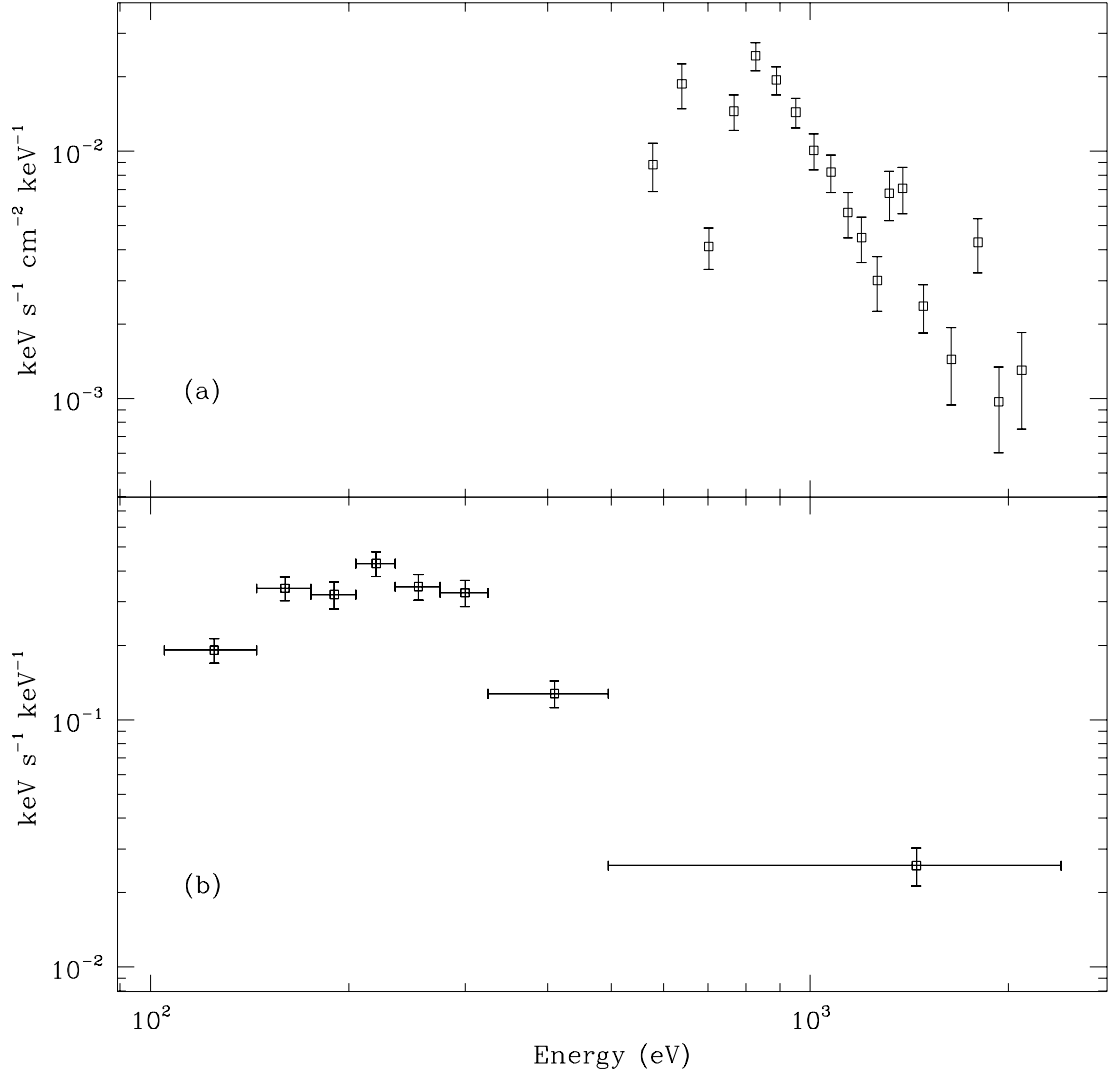}
\caption[]{\bbxrt\ (detector A) spectrum of $\zeta$ Pup (O4 If) taken from
  Corcoran \etal\ (1993) (a).  This spectrum has been ``unfolded''
so as to correct for the energy-dependent sensitivity of the detector.
Note the peak flux is near 0.7 keV, once the oxygen K-edge feature
near this energy is corrected for. 
Given the parameters of $\zeta$ Pup 
($\Mdot = 5 \times 10^6$ \Msunyr; 
$\vinf = 2.2 \times 10^3$ km s$^{-1}$, 
$R_{\ast} = 1.3 \times 10^{12}$ cm)
we find a value of $E_1 \approx 2$ keV, however.  Note also that the power-law indices
of the slopes of the low-energy and high-energy portions of the spectrum differ by at
least six. The \rosat\ PSPC spectrum of
$\lambda$ Sco (B1.5 IV) (b) has a peak energy (corresponding to $E_1$) that is in much better
agreement with the predicted value of 0.1 keV.  The slightly higher observed peak value
is entirely consistent with the finding that the mass-loss rates of B stars have been
systematically underestimated (Cohen et al. 1997).}
\label{Fig:two_stars} 
\end{figure}

Using the simple power-law absorption and emission models, the energy-dependencies in
the two domains have slopes that differ by $a(1-s)$.  Furthermore, if $a(1-s)-b > 0$
then $E_x = E_1$ is also the energy of peak flux, and the value of this peak is dictated
solely by the {\it absorption} properties of the wind.  Given equation (\ref{eqn:e1})
one would expect the observed values of $E_1$ to vary significantly from low-density B
star winds to high-density O star winds.  However, although some variation in the peak
intensities are seen, it is much less dramatic than this model predicts.  In addition,
the relative slopes of the optically thin and optically thick parts of observed X-ray
spectra are generally  seen to differ by much more than $a(1-s)$, given reasonable
values of $a$ and $s$.  These trends are shown in Figure \ref{Fig:two_stars} for a
representative O star (\zpup) and a representative early-type B star ($\lambda$ Sco).

It is thus apparent that a universal power-law emissivity function is too simple an
assumption.  An obvious problem with this functional description is that it implies the
presence of plasma having an arbitrarily high temperature.  It therefore seems
appropriate to consider a truncation to the power-law emissivity introduced in equation
(\ref{eqn:xi_of_ex}) that corresponds roughly to the maximum shock energy in the wind. 
We therefore modify the emissivity as 
\begin{equation}
\xi (E_x) = { \xi_{\ast} \over E_x^b } e^{-E_x/E_{max}},
\label{eqn:xi_of_ex_cutoff}
\end{equation}
where $E_{max}$ is the maximum plasma temperature, corresponding to the maximum shock
energy.  The luminosity spectrum is now given by
\begin{mathletters}
\begin{eqnarray}
{ dL_x  \over dE_x } = L_1 
\left ( { E_x \over E_1 } \right )^{a(1-s)-b} e^{-E_x/E_{max}} ; ~ E_x < E_1~
\label{eqn:lx_of_ex_cut_a}
\\
        = L_1 \left ( { E_x \over E_1 } \right )^{-b} e^{-E_x/E_{max}} ~~~~~~~ ; ~ E_x > E_1 ,
\label{eqn:lx_of_ex_cut_b}
\end{eqnarray}
\end{mathletters}
where $L_1$ has the same definition as before.  The exponential cut-off in this model
brings the luminosity spectrum more in line with the data shown in Figure
\ref{Fig:two_stars}, as well as that for most other OB stars.  

With this general expression for the X-ray spectra, one sees that there are two
different types of spectra that can occur.  If $E_{max} > E_1$ then the high-energy fall
off of the spectrum occurs above $E_{max}$, where the spectrum would be decreasing
as a power-law already.  Relative to the simple power-law expression model, this
spectrum has a peak determined by $E_1$.  Effectively, then, the main
change is to cause a much steeper high-energy fall-off in the spectrum.  
However, if $E_{max} < E_1$ then the peak in the spectrum
is at a softer energy than it otherwise would have been (at an energy dictated by
$E_{max}$ rather than $E_1$), and the fall-off in the spectrum with increasing photon
energy occurs that much sooner.  

When O star X-ray data are fit with two-temperature models (e.g. Hillier et al. 1993),
the higher temperature component tends to be approximately $5 \times 10^6$ K, rarely
exceeding $10^{7}$ K.  Even for the cooler B stars, the highest temperature component
derived from model fitting is roughly $5 \times 10^6$ K (Cassinelli et al. 1994),
indicating that $E_{max}$ should be relatively constant for hot stars, with a value near
$0.5$ keV.  This is in contrast to the values of $E_1$, which vary by more than an order
of magnitude from mid-B to early-O stars.   Therefore, which of these two cases applies
to a given star depends primarily on the value of $E_1$, which is governed largely by
the mass-loss rate.  The case with $E_{max} < E_1$ is thus appropriate to O stars, with
their optically thick winds, while the case with $E_{max} > E_1$ is relevant for B
stars, with their thinner winds.  For these low-density winds, the peak energy in the
spectrum is determined by the value of $E_1$, whereas for the higher density, optically
thick winds the peak of the spectrum is relatively constant at the universal value of
$E_{max} \approx 0.5$ keV.

\subsection{The \lxlb\ Scaling Relation from the Full Spectrum}
\label{sec:windscale_full}

With the parameterization of the energy-dependence of wind attenuation and wind emission
introduced in the previous subsection, we can return to the explanation for the scaling
relationship discussed in \S \ref{sec:windscale}.  Because the O stars and early-B stars
for which the \lxlb\ scaling holds have optically thick winds with $E_{max} < E_1$, the
optically thin portion of the wind ($E_x > E_1$) contributes negligibly to the overall
X-ray luminosity and we therefore can consider only the optically thick expression,
given by equation (\ref{eqn:lx_of_ex_cut_a}).  Furthermore, because of the negligible
contribution of the high-energy portion of the spectrum, we can carry out the
integration over all energies without significantly affecting the result.  The total
X-ray luminosity emergent from a hot star wind in this model is then given by
\begin{mathletters}
\begin{eqnarray}
L_x \approx  { (\xi_{\ast} C_s^2 f_v)_o  \over 2 R_o^s ( 1 - s) } \,
\left ( { 4 \pi R_o \over \kappa_{xo}  } \right )^{1-s} 
\left ( { \Mdot \over \vinf } \right )^{1+s} \int_0^{\infty}  
E_x^{a(1-s)-b} e^{-E_x/E_{max}} \, dE_x 
\label{eqn:lx_int_a}
\\
= { (\xi_{\ast} C_s^2 f_v)_o  \over 2 R_o^s ( 1 - s) } \,
\left ( { 4 \pi R_o \over \kappa_{xo}  } \right )^{1-s} 
\, \left ( { \Mdot \over \vinf } \right )^{1+s} E_{max}^{1+a(1-s)-b} \,
\Gamma(a-as-b).
\label{eqn:lx_int_b}
\end{eqnarray}
\end{mathletters}
This expression preserves the $\lx \sim (\mdot/\vinf)^{1+s}$ dependence which yields
$\lx \sim \lb$ with $s \approx -0.25$ or $s \approx -0.4$, as we derived 
in \S \ref{sec:windscale} for monochromatic X-rays.  
We note that this analysis holds only for the optically thick
winds that correspond to O stars which have $\lxlb \approx 10^{-7}$.

The relationship for monochromatic optically thin winds, $\lx \sim 
(\mdot/\vinf)^2$, is also consistent with the energy-dependent treatment, so long as
$E_1$ is below the soft edge of the instrumental bandpass.  Otherwise the observed
spectrum has significant contributions from both an optically thin power-law
region and an optically thick power-law region. In this hybrid case, the dependence of
the X-ray luminosity on the bolometric luminosity is complicated and falls somewhere
between the linear dependence of the thick winds and the much steeper dependence of the
thin winds.  This would cause a less discontinuous fall-off in the $\lxlb$
relationship than is inferred from the monochromatic analysis in \S \ref{sec:windscale}.

\section{Conclusions} \label{sec:conclusions}
 
Despite its approximate nature, the analysis described here provides a useful initial
framework for examining the interplay among the basic processes of X-ray emission and
absorption that determine the observed X-ray spectra from a hot-star wind.

Some of the main points are: 
 
\begin{enumerate}

\item The simple procedure (cf. eq. [\ref{eqn:exospheric}]) of
including only the X-ray emission above the radius $R_1$ of unit 
optical depth (the exospheric approximation) provides an accurate means to 
account for X-ray attenuation effects.  
A key aspect of this approximation is the 
assumption that $v \approx \vinf$, which is indicated theoretically.  
If the X-rays are formed deeper in the wind, then this constant-velocity 
approximation will underestimate the X-ray luminosity, but  
in Appendix \ref{App:beta} we estimate that the error is generally no more than about 50\%.

\item  For optically thin winds, the X-ray luminosity increases as $(\Mdot /
\vinf)^2$, as would be expected based on the density-squared sensitivity of the thermal
emission processes.  However, the dependence of the X-ray luminosity on wind density is
less steep in the case of optically thick winds, with $L_x \sim (\Mdot / \vinf)^{1+s}$, where the
X-ray filling factor has a power-law radial dependence, $f \sim r^s$.  

\item Indeed, the natural scaling of the x-ray luminosity in hot-star winds is with the
wind density parameter $\mdot/\vinf$, and not with stellar parameters such as the
bolometric luminosity.

\item  However, if one introduces a radial power-law scaling of the filling factor, then
the observed $\lx \sim \lb$ relation can be recovered with a value of $s \approx -0.25$
to $s \approx -0.4$ for the radial power-law index.

\item The energy dependence of the unit-optical-depth radius $R_1$ can
play an important role in the shape of the energy spectrum.  Indeed,
in the simple shock model presented here, the peak of the spectrum
occurs at the energy $E_1$ for which X-rays become optically thin at
the initial shock onset radius $R_o$, as long as the emissivity does 
not begin to fall-off rapidly at energies below $E_1$.  This condition 
should be satisfied for the low-density B star winds.  However, for O star 
winds, the peak of the spectrum is governed by the high-energy cut-off in 
the shock strength distribution. 

\item The X-rays observed by satellite telescopes are, especially for stars with
optically thick winds, only a fraction of the total X-ray production.  A significant
amount of the X-ray production is absorbed by the optically thick winds of these stars. 
In addition, a significant amount of the total generated X-ray emission may fall outside
of the observational bandpasses, especially in the very soft X-ray and EUV.
  
\end{enumerate}

\blankline  
 
Future work will apply the principles illuminated here to interpretation of more
fundamental, numerical simulations of X-ray emission from instability-generated shocks,
to a detailed analysis of shock heating and cooling mechanisms, and to the
interpretation of observational data.  This will help clarify the strengths and
weaknesses of the wind instability paradigm for explaining the observed X-ray spectra
and scaling relations.

\acknowledgements
This research was supported in part by NASA grant NAG5-3530 to the Bartol Research 
Institute at the University  of Delaware.

\appendix

\section{Exospheric Model with a Beta Velocity Law}\label{App:beta}

Although there is good reason to believe that the X-ray production in hot star winds 
occurs primarily in regions of the wind where the velocity is close to the terminal
velocity, it is possible to relax this assumption and still analytically solve
for the X-ray luminosity under the exospheric approximation (eq. [\ref{eqn:exospheric}]),  
if the filling factor is radially constant.

The exospheric approximation is given by
\begin{equation}
L_x \approx 8 \pi^2 \int_{R_1}^\infty  \eta (r) \, r^2 dr ,
\end{equation}
where the emissivity $\eta \sim \rho^2$, and $R_1$ is the radius at which the radial optical depth 
is unity, $\tau(0,R_1) \equiv 1$.
For a `beta' velocity law of the form
\begin{equation}
v(r) = \vinf \left( 1 - R_{\ast}/r \right)^{\beta} ,
\label{eqn:beta}
\end{equation}
the radial optical depth  is given by the integral
\begin{equation}
\tau(0,r) =
{\kappa_x \Mdot \over 4 \pi \vinf } \int_r^\infty { dr' \over r'^2 (1 - R_{\ast}/r')^{\beta}},
\label{eqn:taur_beta}
\end{equation}
which through a variable substitution $y = \left( 1 - R_{\ast}/r \right)$ can be 
readily evaluated as
\begin{equation}
\tau(0,r) = {\kappa_x \Mdot \over 4 \pi \vinf R_{\ast} 
(1 - \beta)}[1 - \left( 1 - R_{\ast}/r \right)^{1-\beta} ] .
\label{eqn:taurbeta2b}
\end{equation}
Setting $\tau=1$ leads to a new expression for the unit-optical-depth radius, 
\begin{equation}
R_1^{beta} = R_{\ast} 
\left[ 1 - \left(1 - {R_{\ast}(1-\beta) \over R_1^0} 
\right)^{1 \over 1 - \beta} \right]^{-1} ,
\label{eqn:r1beta}
\end{equation}
where $ R_1^0 \== \kappa_x \Mdot / 4 \pi \vinf $ is the previous result for a constant
velocity flow, which simply represents the special case $\beta=0$.

Making the same change of variables in the exospheric expression for the X-ray luminosity 
and using the newly defined lower limit of integration, $R_1^{beta}$, one finds
\begin{equation}
L_x^{beta} = {\xi C_s^2 f_v \mdot^2 \over 2 R_{\ast} (1-2\beta) \vinf^2} 
\left[ 1 - \left (1 - {R_{\ast} (1 - \beta) \over R_1^0} \right)^{1 - 2\beta \over 1 - \beta}  \right].
\label{eqn:lxbeta}
\end{equation}
Scaled by the constant-velocity exospheric expression (eq. [\ref{eqn:exospheric}]), 
this can be cast as a correction factor,
\begin{equation}
L_x^{beta} / \lx = {R_1^0 \over R_{\ast} (1 - 2\beta)} \left[ 1 - 
\left (1 - {R_{\ast} (1 - \beta) \over R_1^0} \right)^{1 - 2\beta \over 1 - \beta}  \right].
\label{eqn:lxbolx}
\end{equation}

Note that as $R_1^0 / R_{\ast}$ approaches infinity, $L_x^{beta} / \lx$ approaches unity.  
Furthermore, for $\beta = 0.8$ if $R_1 = 2 R_{\ast}$ then $L_x^{beta}/\lx \approx 1.25$.
Choosing an $R_1^0$ that gives $R_1^{beta} = 1.5 R_{\ast}$, which seems a reasonable
minimum radius for onset of instability-generated X-rays,   
$L_x^{beta}$ is just over 50\% greater than \lx.

\section{Comparison of Power-law Emission Spectra to Numerical Models}\label{App:em}

Given that the optically thin, thermal emission expected from the shock-heated regions
of hot-star winds is dominated by lines ({\it e.g.} Raymond \& Smith 1977; Mewe,
Groenenschild, \& van den Oord 1985), we may ask how well the smooth power-law models we
have assumed in this paper represent the true spectra. We thus created a series of
collisional equilibrium spectral models using the MeKaL code (Mewe, Kaastra, Liedahl
1995) and assuming B-star abundances (Geis \& Lambert 1992).  These are power-law
differential emission models of the form ${dEM \over dT} \sim 1/T^c$ that are
constructed from a superposition of multiple isothermal models. The individual
isothermal models used to make the power-law differential emission models have
temperatures on the interval $5.0 \leq \log T({\rm K}) \leq 7.7$.  The spectra were
calculated on the range 15 eV to 2500  eV.  Although 100 eV is generally taken to be the
lower bound of the X-ray bandpass, significant emission occurs at lower energies, even
for very hot plasma. We have compared these single power-law temperature distribution
models to the single power-law spectral models described by equation
(\ref{eqn:xi_of_ex}).

\begin{figure}
\plotone{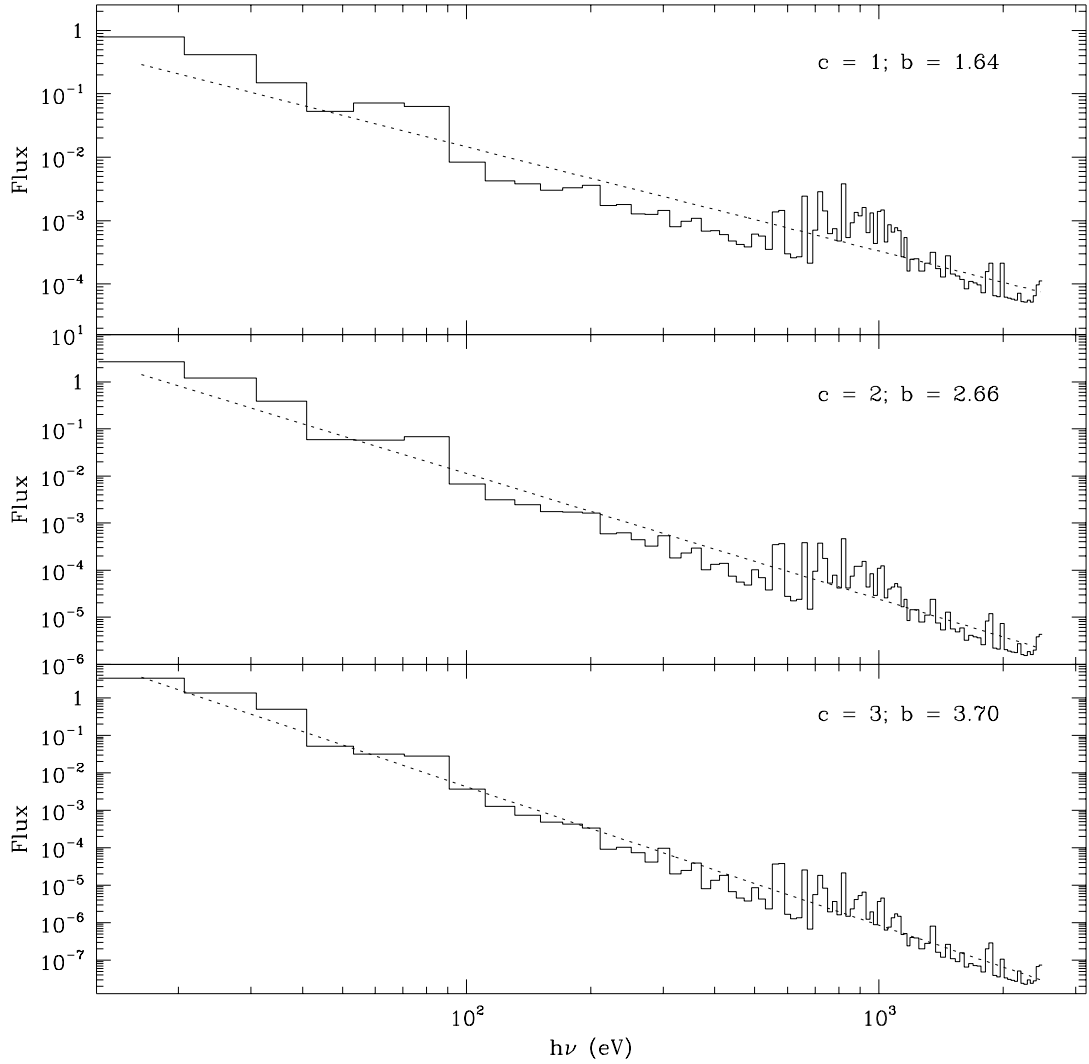}
\caption[]{Temperature power-law spectral models (solid lines) calculated from the MeKaL 
code are shown along with the power-law spectral fits (dotted lines).  
The best-fit power-law spectral indices (b) are indicated along with the differential emission measure slope used in making the models (c).}
\label{Fig:linfit3} 
\end{figure}

Currently, the best spectral resolution available from space-based observatories is $
\lambda/\Delta\lambda  \approx 20$ from {\it BBXRT} and {\it ASCA}.  We therefore binned
the detailed models at slightly better than this resolution, and fit the resulting
spectra with single power-laws, $\xi \sim E_x^{-b}$.  Three examples of this exercise
are shown in Figure \ref{Fig:linfit3}.  The power-law assumption is seen to be quite
good.  Reasonable $\chi^2$ values result from these fits if statistical uncertainties of
only a factor of 2 to 3 are assumed for each bin.  It should be kept in mind that the
actual resolution of most available X-ray data sets is significantly worse than the
binning of these models indicate, so that the observed data are, in fact, even more
well-described by power-laws.

There are several interesting things to note from this exercise.  First, the global
structure in these model spectra are dominated by the emission bump centered near 0.8
keV.  This feature is due to thousands of iron L-shell emission lines.  Second, a very
significant fraction of the flux emerges not in the soft X-ray, but actually in the EUV,
below 100 eV.  Although this radiation is very significant (not least because of the
role it plays in determining the wind excitation and ionization conditions), it has been
directly observed in only one hot star, $\epsilon$ CMa (Cohen et al. 1996).  Finally,
the best-fit spectral power-law slopes are generally steeper than the power-law slopes
of the differential emission measure (plasma temperature) distributions by roughly
$0.7$.  This factor is a reflection of the fact that on the temperature range $10^5 <
{\rm T} < 10^7$, hotter plasma emits less efficiently than cooler plasma.  Indeed, the
slope of the frequency-integrated cooling-curve between the temperatures of $10^5$ and
$10^7$ K (\eg\ Raymond, Cox, \& Smith 1976) is very close to the value $0.7$.

\newpage

\end{document}